\documentclass[preprintnumbers,amsmath,amssymb]{revtex4}

\usepackage{graphicx}
\usepackage{dcolumn}
\usepackage{bm}
\usepackage{rotating}
\usepackage{color}
\usepackage{soul}

\begin{document}
\title{Seeking high temperature superconductors in ambient\\from exemplary beryllium-based alloys}
\date{\today}
\author{X. H. Zheng$^1$}
\email{xhz@qub.ac.uk} 
\author{J. X. Zheng$^2$} 
\affiliation{$^1$Department of Physics, Queen's University of Belfast, BT7 1NN, N.  Ireland}
\affiliation{$^2$Department of Electrical and Electronic Engineering, Imperial College London, SW7 2AZ, England}

\begin{abstract}
With the help of the McMillan formula and virtual crystal model, we predict $T_c$ may exceed 34 K in a beryllium-based alloy with a specific composition, reminiscent of $T_c = 35$~K in the first cuprate superconductor. This may similarly inspire research efforts to seek high temperature superconductors in ambient.\\

\end{abstract}
\maketitle

\section{introduction}
Recently Eremets and colleagues found via experiment that in lanthanum superhydride (LaH$_{10}$) the superconducting transition temperature, $T_c$, can reach 250 K under extreme pressure \cite{Eremets}, following theoretical predictions by Ashcroft in two publications in 1968 and 2004 \cite{Ashcroft1, Ashcroft2}.  Here, with the help of the McMillan formula and virtual crystal approximation, we follow suit to predict that $T_c$ may exceed 34 K in a beryllium-based alloy, to serve as a timely path towards high temperature conventional superconductors in ambient, for two reasons.  First, Bednorz and M\"uller found a similar $T_c$ (35 K) in a cuprate in 1986 \cite{Bednorz}, where the subsequent events have lead to a rapid increase in $T_c$ in cuprates.  Second, in contrast to the largely empirical cuprate research, our prediction stems from a solid theoretical foundation, describing a detailed relation between $T_c$ and alloy composition, ready for experimental verification to facilitate future advancement.

\section{McMillan formula}
William L.~McMillan, late professor of physics at the University of Illinois, Urbana-Champaig, solved the Eliashberg equations numerically via iteration, with a number of simplifications \cite{McMillan}. The equations are linearised at $T = T_c$, and the solutions are assumed to have just two values, $\Delta_0$ and $\Delta_\infty$, defined immediately beneath and deeply inside the Fermi surface, respectively.  In addition the electron-phonon spectral density, $\alpha^2F(\omega)$, is assumed to be a product of $\alpha$ and $F(\omega)$, where $\alpha$ is a constant and $F(\omega)$ the phonon density of states (from Nb neutron scattering experiments for any bcc lattice, assumed to vanish for $\omega < 100$ K). Over the course of iterations, $T_c$ and the Coulomb pseudopotential, $\mu^*$, are kept constant, and $\alpha$ adjusted continuously to
keep $\Delta_0$ constant. The formula
\begin{eqnarray}\label{eq:1}
T_c = \frac{\Theta}{1.45}\exp\left[-\frac{1.04(1 + \lambda)}{\lambda - \mu^*(1 + 0.62\lambda)}\right]
\end{eqnarray}
results from numerical fitting, where $\Theta$ stands for the Debye temperature or the Bloch-Gr\"uneisen characteristic temperature \cite{Ziman}, whenever appropriate, and $\lambda$ the electron-phonon coupling factor.  Sample outputs of Eq.~(\ref{eq:1}), $\mu^* = 0.13$, are shown as open circles in FIG.~1.

\begin{figure}[h]
\centering\includegraphics[width=8cm]{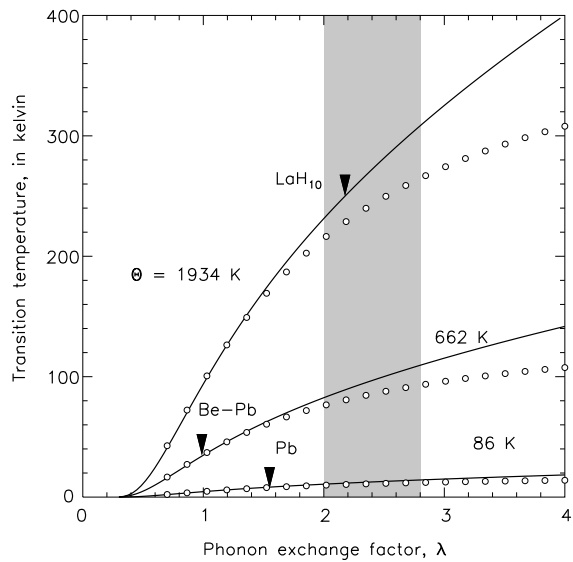}
\caption{$T_c$ against $\lambda$ from Eqs.~(1) (open circles) and (\ref{eq:3}) (curves). The grey area marks the range $2\leq\lambda\leq2.8$, and downward arrowheads indicate $T_c$ and $\lambda$ in Pb ($\mu^* = 0.13$), Be$_{43}$Pb$_{57}$ ($\mu^* = 0.116$) and LaH$_{10}$ ($\mu^* = 0.13$).}
\label{fig:1}
\end{figure}

There has been a long-standing myth that McMillan ``extrapolated his equation for $T_c$ beyond its regime of validity to fortify claims that 30 K would be the upper limit for electron-phonon coupling" \cite{Pickett}. We wish to make it clear that McMillan did not apply his formula beyond its regime of validity, nor did he fortify any claim to limit $T_c$ in general.  Conversely it was McMillan himself who tried hard to establish a regime of validity on Eq.~(\ref{eq:1}), bound at $\lambda = 2$, to define maximum $T_c$ against $\Theta$. McMillan allowed $\Theta$ to ascend and imposed no upper limit on $T_c$.

Letting $\lambda\rightarrow\infty$ and $\mu^* = 0.13$, we find from Eq.~(\ref{eq:1}) an estimation $T_c = \Theta/4.49$, which McMillan apparently considered to be too optimistic.  He argued that $\Theta\propto\langle\omega\rangle$ and $\lambda\propto1/\langle\omega^2\rangle$, so that $T_c$ from Eq.~(\ref{eq:1}) (with the constant 1.04 and $\mu^*$ replaced by 1 and 0 respectively for simplicity) declines when on average $\omega$ is either too large or too small.  Letting $dT_c/d\langle\omega\rangle = 0$, McMillan found $\lambda = 2$ as the condition to maximize $T_c$ \cite{McMillan}, giving via Eq.~(\ref{eq:1}) $T_c = \Theta/9.00$ as a refined estimation (constant 1.04 reinstated, $\mu^* = 0.13$).

McMillan did however allow $\lambda$ to exceed 2, in order to squeeze higher $T_c$ out of Eq.~(\ref{eq:1}).  He divided superconductors into classes, with common $\Theta$ and $\mu^*$, and placed a realistic example, with a certain $\lambda$ and $T_c$, into each class. The value of $\lambda$ was then raised to 2.8,  amounting to an estimation $T_c = \Theta/7.30$ ($\mu^* = 0.13$).  In response Eq.~(\ref{eq:1}) produced a higher value of $T_c$, which McMillan claimed to be the maximum in each class.  By doing so he claimed 9.2, 22, 28 and 40 K would be maximum $T_c$ in superconductor classes exemplified by Pb, Nb, Nb$_3$Sn and V$_3$Si~\cite{McMillan}.

The McMillan $T_c$ from Eq.~(\ref{eq:1}) can be compared with another estimation, $T_c = 0.182\;\bar\omega_2\sqrt{\lambda}$, from the Eliashberg equations in the Matsubara representation \cite{Allen}.  It can be written as $T_c = \Theta\sqrt{\lambda}/7.8$ because $\bar\omega_2 = \Theta/\sqrt{2}$ when Debye phonons are applied, see Appendix A.  It was commented that the Matsubara $T_c$ does not imply a restriction of the type $T_c < \Theta$ but can {\it in principle} lead to an {\it arbitrarily large} $T_c$ \cite{Allen}, which also implies an arbitrarily large value of $\lambda$.  However, since $\lambda$ measures the strength of the electron-phonon interactions, it cannot grow indefinitely.  In addition, at $T = T_c$, the real part of the superconducting energy gap function, $\Delta_r(\omega)$, is involved in just one of the Eliashberg equations in the form 
\begin{equation}\label{eq:2}
\Delta_r(\omega) = \lambda\int_0^\infty K(\omega, \omega')\Delta_r(\omega')d\omega'
\end{equation}
where $K(\omega, \omega')$ is the real part of the Eliashberg kernel (renormalisation function included) \cite{McMillan} with $\alpha^2F(\omega')$ being normalized by $\lambda$ (amplitude = 1 in the case of Debye phonons, see Appendix A).  The imaginary part of the gap function can be found from a similar equation.  Eq.~(\ref{eq:2}) is a Fredholm integral equation of the second type, where $\lambda$ is well known to have an upper limit \cite{Crushing}, although we do not know exactly how large $\lambda$ can be.

\section{Allen-Dynes modification}
Allen and Dynes pointed out that $T_c$ from Eq.~(\ref{eq:1}) can be highly accurate for all known materials with $\lambda < 1.5$ but in error with larger values of $\lambda$ \cite{Allen}.  To solve the problem they modified Eq.~(\ref{eq:1}) and found
\begin{eqnarray}\label{eq:3}
T_c = \frac{\Theta}{1.56}\exp\left[-\frac{1.04(1 + \lambda)}{\lambda - \mu^*(1 + 0.62\lambda)}\right]f_1f_2
\end{eqnarray}
where $f_1f_2\sim1$ when $\lambda$ is small, reducing Eq.~(\ref{eq:3}) back to the McMillan formula, $f_1f_2\propto\lambda^{1/2}$ when $\lambda$ is large to boost the value of $T_c$.  In its original form Eq.~(\ref{eq:3}) includes a coefficient $\omega_{\mbox{\scriptsize log}}$ \cite{Allen}.  We assume Debye phonons to replace $\omega_{\mbox{\scriptsize log}}$ with $\Theta$ for a clear comparison between Eqs.~(\ref{eq:1}) and (\ref{eq:3}), see Appendix A.

In FIG.~1 we show the output of Eq.~(\ref{eq:3}) as the curves, which are indeed noticeably higher than the output from Eq.~(\ref{eq:1}) when $\lambda > 1.5$.  We use the grey area to mark the range $2\le\lambda\le2.8$ to show how far McMillan applied Eq.~(\ref{eq:1}) beyond his own regime of applicability.  Letting $\mu^* = 0.13$ and $\Theta = 86$ K, we find $T_c = 9.6$ and 11.8 K from Eq.~(\ref{eq:1}) when $\lambda = 2.0$ and 2.8, respectively, with a moderate increase (23\%) in $T_c$. In comparison, with the same values of $\mu^*$, $\Theta$ and $\lambda$, we find from Eq.~(\ref{eq:3}) $T_c = 10.2$ and 13.6 K, with a slightly higher increase (33\%).

\begin{figure}[h]
\centering\includegraphics[width=8cm]{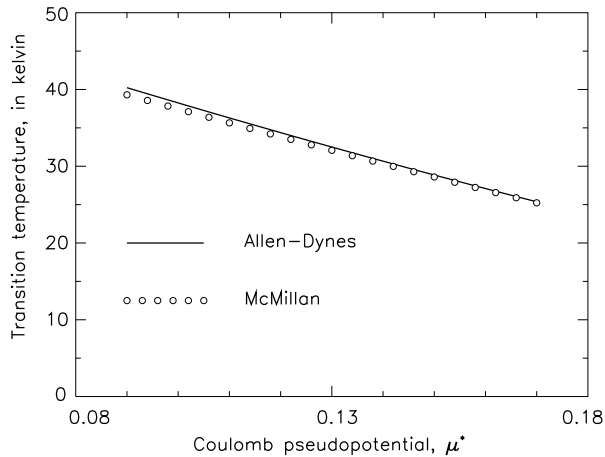}
\caption{$T_c$ against $\mu^*$ from Eqs.~(1) (circles) and (\ref{eq:3}) (curve),  with $\Theta = 662$ K and $\lambda = 0.988$ for Be$_{43}$Pb$_{57}$.}
\label{fig:1}
\end{figure}

So far we have assumed $\mu^* = 0.13$ or $\sim$0.12, close to the experimental average \cite{McMillan}.  In practice $\mu^*$ is extracted from tunnelling data in conjunction with a numerical procedure \cite{McMillan2}, where $\mu^*$ has to be adjusted continuously and carefully, in order to match the experimental and numerical values of the edge of the energy gap function, because the numerical gap edge is sensitive to the value of $\mu^*$.  However, with $\Theta = 662$ K and $\lambda\sim1$ for a Be-Pb alloy and over the range $0.09\leq\mu^*\leq0.17$, $T_c$ from Eq.~(\ref{eq:1}) varies from 39.3 to 25.2 K (36\% reduction), whereas $T_c$ from Eq.~(\ref{eq:3}) varies from  40.2 to 25.4 K (37\% reduction), see FIG.~2. In other words, $\mu^*$ may not have exceedingly strong effect on $T_c$ from either the original McMillan formula, or its Allen-Dynes modification.

\section{virtual crystal approximation}
On account of our discussions in the previous two sections, scrutinizing the rather moderate effects of $\lambda$ and $\mu^*$ on $T_c$ from both Eqs.~(\ref{eq:1}) and (\ref{eq:3}), it has become clear that we have little choice but to seek materials with high $\Theta$ if we wish to raise $T_c$ to above, say, 200 K.  Very recently Eremets and colleagues found $T_c\simeq250$~K in LaH$_{10}$ under high pressure \cite{Eremets}.  From the temperature curve of resistance over $250\leq T\leq273$ K in sample R/9 in FIG.~1 in \cite{Eremets}, we find $\Theta\simeq1934$ K via the Bloch-Gr\"uneisen formula \cite{Ziman}, which means we must have $\lambda = 2.39$ in Eq.~(\ref{eq:3}), compared with $\lambda = 2.52$ in Eq.~(\ref{eq:1}), $\mu^* = 0.13$ in both cases.  It is apparent that it is impossible to achieve $T_c = 250$ K when $\Theta = 86$ or 662 K, at least with $\lambda\leq4$.

The Debye temperature, $\Theta$, is the temperature of a crystal's highest normal mode of vibration, and it correlates the elastic properties with the thermodynamic properties such as phonons, thermal expansion, thermal conductivity, specific heat and lattice enthalpy \cite{Low}.  In general a material is harder and more brittle the higher its $\Theta$.  While details are beyond the scope of our discussion, we wish to mention that $\Theta = 1440$ and 2230~K in beryllium and carbon respectively, far exceed other elements in the periodic table \cite{Kittel}, and many high $\Theta$ compounds also contain carbon~\cite{Low}.  Here we investigate $T_c$ in Be-based alloys, Be-Pb alloys in particular, where Be has extremely high $\Theta$ but low $T_c$ (0.026 K), whereas Pb has very low $\Theta$ ($\sim$86 K from the Bloch-Gr\"uneisen formula) \cite{Ziman} but one of the highest $T_c$ (7.19~K) among simple metals, hoping to gain an advantage from both ingredients.

We model metallic alloys with the virtual crystal approximation, which has been employed extensively to study for example the band structure of disordered alloys \cite{Bellaiche}.  In the approximation some symmetry and periodicity are assumed for the lattice, composed by fictitious or `virtual' atoms that interpolate between the behaviour of the atoms in the parent materials \cite{Bellaiche}.  Therefore, with respect to $T_c$ prediction in, say, Be-Pb alloys, we need only the phonon spectra in Be and Pb for the parent values of $\Theta$, $\lambda$ and $\mu^*$, which have been well investigated \cite{Allen, Mitrovic}, in order to determine $\Theta$, $\lambda$ and $\mu^*$ in each offspring alloy.

We first test the virtual crystal approximation on Tl-Pb alloys against experiment.  In Eqs.~(\ref{eq:1}) and (\ref{eq:3}) we let
\begin{equation}\label{eq:4}
\begin{array}{ll}
\Theta = x\Theta_1 + (1 - x)\Theta_2,\\
\\
\lambda = x\lambda_1 + (1 - x)\lambda_2\\
\\
\mu^* = x\mu_1^* + (1 - x)\mu_2^*
\end{array}
\end{equation}
in accordance with the usual practice of virtual crystal approximation, over $0\leq x\leq1$ \cite{Bellaiche}.  In Pb we have $\Theta_1 \cong 86$ K (Bloch-Gr\"uneisen temperature) \cite{Ziman}, $\lambda_1 = 1.55$ and $\mu_1^* = 0.13$ \cite{Mitrovic}, giving via Eq.~(\ref{eq:1}) $T_c = 7.65$ K (7.19 K experimentally).  In Tl we have $\Theta_2 = 78.5$ K \cite{Kittel}, $\lambda_2 = 0.795$ \cite{Mitrovic} and let $\mu_2^* = 0.1334$ to balance Eq.~(\ref{eq:1}) for $T_c = 2.36$ K.  We find $T_c$ in the alloy from Eqs.~(\ref{eq:1}) and (\ref{eq:4}) with reasonable accuracy (1.8\% r.m.s.~deviation) against experimental data \cite{Mitrovic} shown as filled squares in FIG.~3.  We also find $T_c$ from Eqs.~(\ref{eq:3}) and (\ref{eq:4}) shown as the line (with slight curvature) crossing the filled squares.

\begin{figure}[h]
\centering\includegraphics[width=8cm]{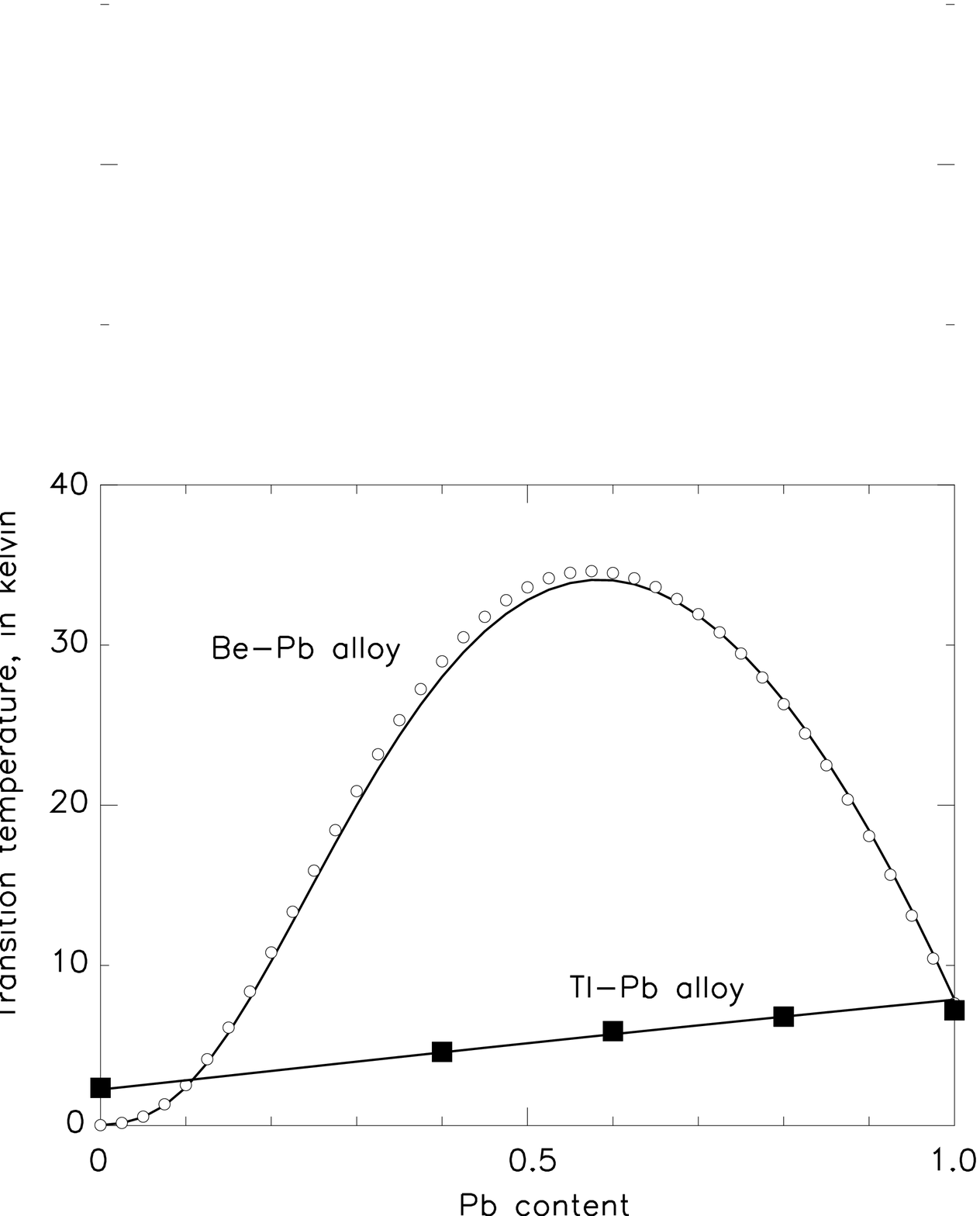}
\caption{Experimental and virtual crystal $T_c$ in Tl-Pb and Be-Pb alloys.  The filled squares are experimental results~\cite{Mitrovic}. Open circles and curves are from Eq.~(\ref{eq:1}) and (\ref{eq:3}), respectively, both applied with  the virtual crystal model in Eq.~(\ref{eq:4}).}
\label{fig:2}
\end{figure}

Next we make a prediction for $T_c$ in Be-Pb alloys with the virtual crystal approximation by directly replacing Tl with Be, where we have $\Theta_2 = 1440$ K \cite{Kittel}, $\lambda_2 = 0.23$ \cite{McMillan} and let $\mu_2^*= 0.09517$, to balance Eq.~(\ref{eq:1}) for $T_c = 0.026$ K.  We plot the outcome of Eqs.~(\ref{eq:1}) and (\ref{eq:4}) as circles in FIG.~3, where $T_c$ reaches a maximum of 34.6~K when $x = 0.57$, $\Theta = 662$ K, $\lambda = 0.988$ and $\mu^* = 0.116$.  We also plot the outcome of Eqs.~(\ref{eq:3}) and (\ref{eq:4}) as the curve close to the circles.  The effect of $\mu^*$ on $T_c$ in Be$_{43}$Pb$_{57}$ can be appreciated from FIG.~2.

\begin{figure}[h]
\begin{minipage}{8cm}
TABLE I: VIRTUAL CRYSTAL DATA
\vspace{2mm}
\begin{ruledtabular}
\begin{tabular}{cccccc}
Crystal&Composition&$\Theta$&$\lambda$&$\mu^*$&$T_c$\footnotemark[1]\\\hline
Be-Al&24:976&452&0.427&0.127&1.18 (1.18)\\
Be-In&31:69&521&0.627&0.120&9.10 (3.40)\\
Be-Nb&23:77&546&0.684&0.117&12.7 (9.21)\\
Be-Pb&43:57&662&0.988&0.116&34.6 (7.19)\\
Be-Bi-Pb&43:20:37&671&1.015&0.115&36.8 (8.95)
\end{tabular}
\end{ruledtabular}
\footnotetext[1]{Experimental $T_c$ in doping metal or alloy \cite{Mitrovic} bracketed.}
\end{minipage}\hfill
\end{figure}

In addition we apply a virtual crystal approximation to make predictions for $T_c$ in other Be-based alloys.  In Al we have $\Theta_1 = 428$ K \cite{Kittel}, $\lambda_1 = 0.432$ \cite{Mitrovic} and let $\mu^*_1 = 0.128$ to balance Eq.~(\ref{eq:1}) for $T_c = 1.18$ K, and find via Eqs.~(\ref{eq:1}) and (\ref{eq:3}) the data in TABLE~I in the Be-Al alloy optimised for maximum $T_c$.  For In we have $\Theta_1 = 108$ K \cite{Kittel}, $\lambda_1 = 0.805$ \cite{Mitrovic} and let $\mu^*_1 = 0.131$ to balance Eq.~(\ref{eq:1}) for $T_c = 3.4$ K, and find the data in TABLE~I in the optimised Be-In alloy.  In Nb we have $\Theta_1 = 275$ K \cite{Kittel}, $\lambda_1 = 0.82$ \cite{McMillan}, and let $\mu^*_1 = 0.124$ to balance Eq.~(\ref{eq:1}) for $T_c = 9.5$ K, to find the data in TABLE I.  Furthermore, viewing Bi$_{35}$Pb$_{65}$ as a single component in the alloy, we interpolate Debye temperatures in Bi and Pb \cite{Kittel} to find $\Theta_1 = 97.55$~K.  Assuming $\mu^*_1 = 0.13$ and $\lambda_1 = 1.60$ to balance Eq.~(\ref{eq:1}) for $T_c = 8.95$ K \cite{Mitrovic}, we predict that a Be-Bi-Pb alloy may reach a maximum $T_c=36.7$~K, also shown in TABLE I.   

\section{previous beryllium alloy experiments}
Superconductivity in Be was investigated actively in both theory and experiment in the 1960-80s, see \cite{Takei} and the references therein.  In one line of enquiry Be vapour is quenched onto cryogenic substrates and cooled to the temperature of liquid helium to form thin films, where $T_c$ was found to reach about 9 K, compared with 0.026 K in bulk Be.  This high $T_c$ phase of Be changes irreversibly to a different phase, with significantly reduced $T_c$, once the substrate of the film has been heated to several tens of kelvin.  In 1985, Takei, Nakamura and Maeda found that, when the Be film is grown on room temperature substrates using the ion beam sputtering technique, $T_c$ in the film can reach $\sim$6 K \cite{Takei}.

In another line of enquiry Be is co-evaporated with other elements to form alloy films on cryogenic substrates. When doped with C, B, W, La, Pd and Ge, values of $T_c$ in the Be-based films always become lower with increasing concentration of the second element.  The authors admitted that they were unable to give a detailed explanation for the systematic depression of $T_c$ \cite{Klein}.

With the virtual crystal model, the declining $T_c$ in, for example, the Be-W films can be readily explained.  In W ($T_c = 0.012$ K) we have $\Theta_1 = 400$ K \cite{Kittel}.  We assume $\mu^*_1 = 0.13$ and let $\lambda_1=1.599$ to balance Eq.~(\ref{eq:1}).  On the other hand, we may have $\Theta_2 = 1440$ K and $\mu^*_2 = 0.0952$ in Be in both bulk and quenched film forms.  We let $\lambda_2 = 0.439$ to balance Eq.~(\ref{eq:1}) for $T_c = 9$ K, and find from Eqs.~(\ref{eq:1}) and (\ref{eq:3}) that $T_c$ indeed declines monotonically from 9 K to nearly zero when $x$ increases from 0 to 1.  Experimentally, $T_c$ declines a lot quicker with increasing $x$~\cite{Klein} likely because Be is no longer in the the $T_c = 9$~K phase when it is diluted too much by W.  

\section{conclusions}
In summary, McMillan could have brought a huge impact to the field of conventional superconductivity in 1968, had he searched high $T_c$ in alloys with high Debye temperatures.  We find $T_c$ may exceed 34 K in Be-based alloys, which is comparable with the first cuprate ($T_c = 35$ K) discovered in 1986 in terms of $T_c$. Our prediction also enjoys the merit that its properties stem from a well-established theoretical foundation, in the shape of the McMillan formula (and its Allen-Dynes modification) and virtual crystal approximation.  Indeed we are able to predict a detailed relation between $T_c$ and alloy content, shown as the open circles in FIG.~3, ready for experimental verification, and bringing with it the insight to facilitate future advancement.

\appendix
\section{}
We adopt the view \cite{McMillan} that $\alpha^2F(\omega)$ is a product of a constant, $\alpha$, and phonon density of states, $F(\omega)$, which in the Debye model is proportional to $\omega^2$ with $\omega\leq\Theta$, so that $\alpha^2F(\omega) = \lambda(\omega/\Theta)^2$ holds if $\omega\leq\Theta$, otherwise it vanishes, to let $\lambda$ be the outcome of integration of $2\alpha^2F(\omega)/\omega$ over $\omega$, as is expected.  By definition we have
\begin{equation}\label{A1}
\omega_{\mbox{\scriptsize log}} = \Theta\exp\left[\frac{2}{\lambda}\int_0^\infty\frac{d\omega}{\omega}\;\alpha^2F(\omega)\ln\left(\frac{\omega}{\Theta}\right)\right]
\end{equation}
where we insert $\Theta$ and $\ln(\omega/\Theta)$ to replace $\ln(\omega)$ in Eq.~(30) in \cite{Allen}, in order to make the argument of the logarithm function a pure number, giving $\omega_{\mbox{\scriptsize log}} = \Theta/\sqrt{e}$ . Similarly we have
\begin{equation}\label{A2}
\bar\omega_2 = \Theta\left[\frac{2}{\lambda}\int_0^\infty\frac{d\omega}{\omega}\;\alpha^2F(\omega)\left(\frac{\omega}{\Theta}\right)^2\right]^{1/2}
\end{equation}
giving $\bar\omega_2 = \Theta/\sqrt{2}$. Thenceforth we find
\begin{eqnarray}
f_1 = \left[1 + \frac{\lambda^{3/2}}{(2.46 + 9.35\mu^*)^{3/2}}\right]^{1/3}
\end{eqnarray}
and
\begin{eqnarray}
f_2 = 1 + \frac{0.17\lambda^2}{\lambda^2 + 4.5(1 + 6.3\mu^*)^2}
\end{eqnarray}
to specify Eqs.~(35) and (36) in \cite{Allen} and hence Eq.~(\ref{eq:3}) here.

\section*{Acknowledgement}
The authors thank the late professor D.~George Walmsley for numerous discussions and inspirations.


\begin{thebibliography}{9}
\bibitem{Eremets}
A. P. Drozdov, P. P. Kong, V. S. Minkov, S. P. Besedin, M. A. Kuzovnikov, S. Mozaffari, L. Balicas, D. Graf, V. B. Prakapenka, E. Greenberg, D. A. Knyazev, M. Tkacz,
and M. I. Eremets, arXiv:1812.01561.
\bibitem{Ashcroft1}
N. W. Ashcroft, Phys. Rev. Lett. 21 (1968) 1748-1749.
\bibitem{Ashcroft2}
N. W. Ashcroft, Phys. Rev. Lett. 92 (2004) 187002.
\bibitem{Bednorz}
J. G. Bednorz and K. A. M\"uller, Z. Phys. B 64 (1986) 189-193.
\bibitem{McMillan} 
W. L. McMillan, Phys. Rev. 167 (1986) 331-344.
\bibitem{Ziman}
J. M. Ziman, Electrons and phonons, Clarendon, Oxford, 2001.
\bibitem{Pickett}
W. Pickett and M. Eremets, Phys. Today May (2019) 52-58.
\bibitem{Allen}
P. B. Allen and R. C. Dynes, Phys. Rev. B 12 (1975) 905-922.
\bibitem{Crushing}
J. T. Crushing, Applied analytical mathematics for physical scientists, John Wiley, 1975.
\bibitem{McMillan2}
W. L. McMillan and J. M. Rowell, Phys. Rev. Lett. 14 (1965) 108-112.
\bibitem{Low}
I. M. Low (ed), Advances in science and technology of M$_{n+1}$AX$_n$ system, Woodhead, Oxford, 2012.
\bibitem{Kittel}
C. Kittel, Introduction to solid state physics, John Wiley, New York, 1986.

\bibitem{Bellaiche}
L. Bellaiche and D. Vanderbilt, arXiv:cond-mat/9908364v1 [cond-mat.mtrl-sci].
\bibitem{Mitrovic}
B. Mitrovi\'c, H. G. Zarate, J. P. Carbotte, Phys. Rev. B 29 (1984) 184-190.
\bibitem{Takei}
K. Takei, K, Nakamura and Y. Maeda, J. Appl. Phys. 57 (1985) 5093-5094.
\bibitem{Klein}
J. Klein, A. L\'eger, S. de Cheveig\'ne, D. MacBride, C. Guinet, M. Belin, D. Defourneau, Solid state Commun. 33 (1980) 1091-1095.
\end{thebibliography}
\end{document}